\documentclass[10pt]{iopart}

\usepackage{graphicx,cite,xcolor,float,iopams}
\expandafter\let\csname equation*\endcsname\relax
\expandafter\let\csname endequation*\endcsname\relax
\usepackage[pagebackref=false,colorlinks,linkcolor=blue,citecolor=blue,urlcolor=blue]{hyperref}
\usepackage{amsmath,amssymb}
\usepackage{blkarray}

\begin{document}
    
\title{Stabilization and Helicity Control of Hybrid Magnetic Skyrmion}

\author{Muzakkiy~P.~M.~Akhir$^1$, Edi Suprayoga$^1$, and Adam~B.~Cahaya$^{2,1,*}$}
\address{$^1$ Research Center for Quantum Physics, National Research and Innovation Agency, Tangerang Selatan 15314, Indonesia}
\address{$^2$ Department of Physics, Faculty of Mathematics and Natural Sciences, Universitas Indonesia, Depok 16424, Indonesia}
\ead{adam@sci.ui.ac.id}

\vspace{10pt}
\begin{indented}
\item[]
\end{indented}

\begin{abstract}
The hybrid skyrmion, a type of magnetic skyrmion with intermediate helicity between Bloch and N\'eel skyrmion, has gained more attraction. It is tolerant toward the skyrmion Hall effect and a potential candidate for quantum bits. We investigated the stabilization and helicity control of the hybrid skyrmion in a two-dimensional magnetic system using an analytical model and micromagnetic simulation. We look at the interplaying factors of the bulk ($D_b$) and interfacial ($D_i$) Dzyaloshinskii-Moriya interactions (DMI) along with the dipolar interaction. We show that the hybrid skyrmion can stabilize through the interplay between interfacial DMI and either bulk DMI or dipolar interaction. We can also control the helicity of the hybrid skyrmion by tuning the ratio of $D_i/D_b$ when there is no dipolar interaction, or simply by adjusting the $D_i$ when the $D_b$ is absent. Our results suggest that hybrid skyrmions can exist within $0 < |D_i| < 0.4$ mJ/m$^2$ for Co-based magnetic systems.
\end{abstract}

\ioptwocol

\section{\label{sec:intro} Introduction}

Magnetic skyrmions (later defined as skyrmion) are a type of magnetic quasiparticle with a nanoscale swirling vortex-like spin texture \cite{Tokura2020}. They are topologically protected, therefore, they are robust and resistant to external disturbances \cite{hang2020, barker2016}. Furthermore, they can be driven by low electric currents or gate voltage \cite{helene2022, sun2021}, thermal gradients \cite{Raimondo2022}, anisotropy gradients \cite{Tomasello2018}, strain gradients \cite{Yanes2019}, and surface acoustic waves \cite{Nepal2018}. Therefore, they could be promising candidates for data storage, spintronics devices \cite{Nagaosa2013, Fert2013}, memory devices \cite{Tomasello2014}, logic gates \cite{Fattouhi2021, Chauwin2019, Zhang2015}, neuromorphic computing \cite{Das2023, Song2020}, and microwave detectors \cite{Finocchio2015}.

Skyrmions were initially found in bulk systems, such as MnSi \cite{muhlbauer2009}, FeCoSi \cite{yu2010}, Cu$_2$OSeO$_3$ \cite{seki2012}, and CoZnMn \cite{tokunaga2015}. However, these skyrmions were only observed at low temperatures and therefore technologically hard to implement. Later, skyrmions were discovered in thin multilayered systems, such as Pt/Co/Ta and Pt/CoFeB/MgO \cite{woo2016}, Ir/Co/Pt \cite{moreau2016}, and Ir/Fe/Co/Pt \cite{soumyanarayanan2017}, which are more favorable for practical applications, as they are easier to stabilize and manipulate at room temperature \cite{Dohi2022, Fert2017}.

Based on its helicity, there are two main types of skyrmions; Bloch and N\'eel skyrmions. The Bloch skyrmions have spins that rotate in a plane perpendicular to the skyrmion's axis, either clockwise or anticlockwise, with a helicity of $\pi/2$. This type of skyrmion is typically found in bulk samples \cite{muhlbauer2009, yu2010, seki2012}. N\'eel skyrmions, on the other hand, have spins that point radially either inward or outward from the center with a helicity of $0$ or $\pi$. This type of skyrmion can be found in thin multilayered systems \cite{woo2016, moreau2016, soumyanarayanan2017}.

Recently, a type of hybrid skyrmion with intermediate helicity between N\'eel and Bloch skyrmion has received more attention \cite{Vakili2020, Kim2018, Wu2017}. Interestingly, it has improved mobility and reduced the skyrmion-Hall effect (SkHE), which prevents the skyrmions from moving parallel to the current flow \cite{chen2017, jiang2017}. It is one among other methods to overcome SkHE such as antiferromagnetic skyrmion-based spin-torque nano oscillator \cite{shen2019,zhou2020} and antiferromagnetic bilayer-skyrmion \cite{zhang2016}. Furthermore, since the hybrid skyrmion can be considered as a superposition of the N\'eel and Bloch skyrmion, it is also a potential candidate as building blocks for quantum bits (qubits) in quantum computers, where information can be stored by utilizing the helicity degree of freedom \cite{Ezawa2023, Psaroudaki2021, Bindal2021}. Therefore, it is important to know the components that can stabilize and control the helicity in hybrid skyrmions.

In most systems, the stability of the skyrmion is affected by the Dzyaloshinskii-Moriya interactions (DMI) \cite{D, M}, which arise from the exchange interaction with spin-orbit coupling and the lack of structural inversion symmetry. There are two kinds of DMI; bulk and interface DMI, which govern the Bloch skyrmions in bulk samples and N\'eel skyrmions in multilayered systems, respectively. The DMI can be described by local and nonlocal electron models \cite{PhysRevB.106.L100408}. In metallic systems, the conduction electron acts as a non-local electron that mediates the indirect exchange interaction with its resultant known as the Ruderman-Kittel-Kasuya-Yosida (RKKY) interaction \cite{RK,K,Y}. In a two-dimensional magnetic system, these can give rise to a magnetic skyrmion texture \cite{Bogdanov2021, hallal2021, Shawn2020}. In addition to these interactions, the dipolar interaction also plays an important role in stabilizing the skyrmion \cite{bernand2020}. 

In this work, we systematically investigate the key components for the stability and helicity control of hybrid skyrmion through an analytical model. We analyze the interplaying factors of interfacial DMI, bulk DMI, and dipolar interaction. 
To validate our findings, we conducted micromagnetic simulations using a Co-based magnetic structure, which is the most common and promising candidate for room-temperature skyrmionic devices.

\section{\label{sec:theo}  Landau - Lifshitz - Gilbert equation in spherical coordinates}

\begin{figure}
    \centering
    \includegraphics[width=6.7cm]{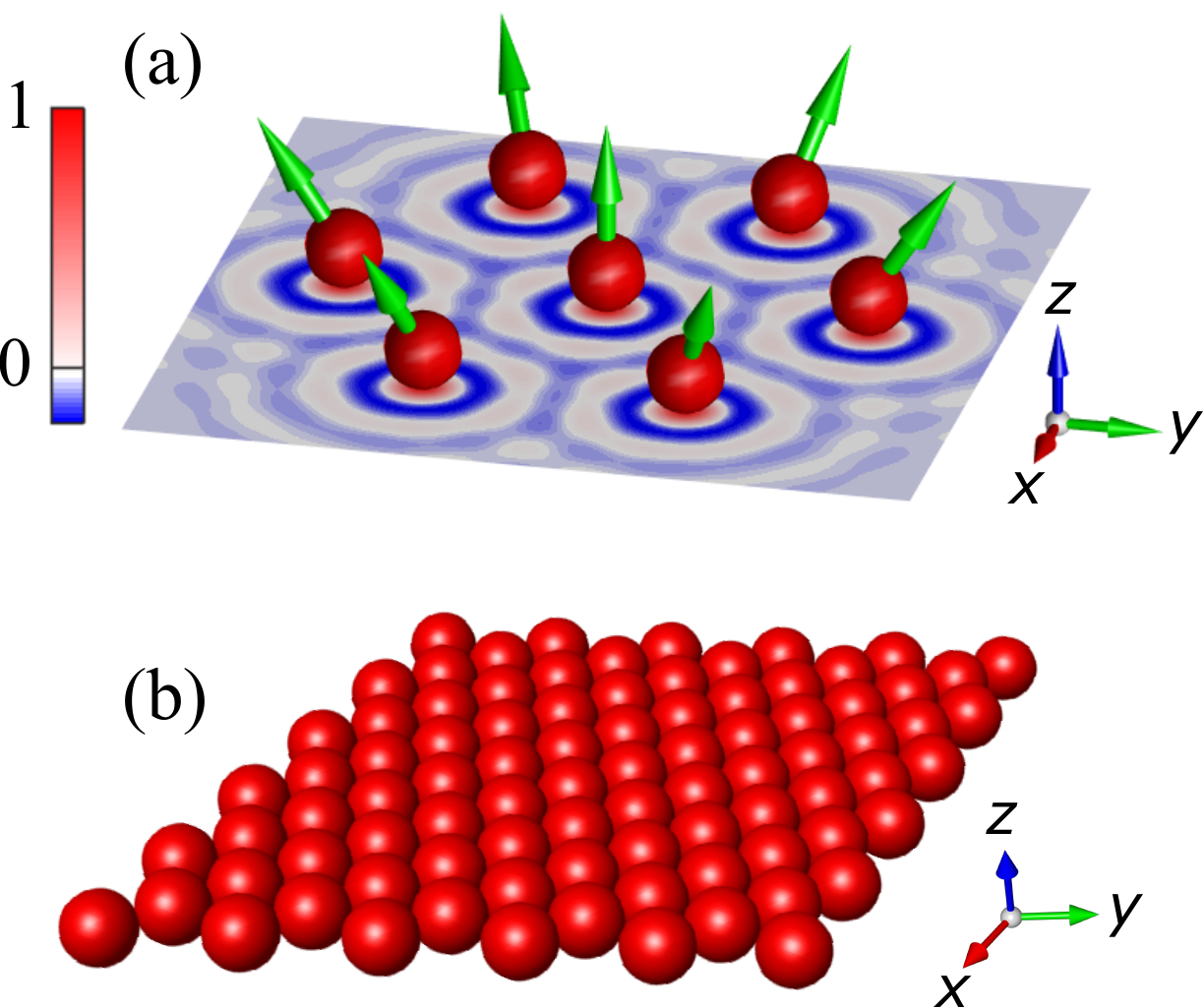}
    \caption{(a) Schematic view of the indirect interactions between magnetic moments in a hexagonal arrangement. The colorbar indicates the exchange strength. (b) Array of magnetic atoms showing the objects arranged in regular pattern. The micromagnetic regime focuses on continuum limit of magnetic moments because the length scale is much larger than the distances between atoms.}
    \label{FigDrawing}
\end{figure}

The two-dimensional magnetic skyrmion can be found at the interface of ferromagnetic metal and nonmagnetic heavy metal, such as Co$|$Pd \cite{Shawn2020}. Microscopically, the skyrmion texture arise from noncollinear direction of nearby spins, as illustrated in Fig.~\ref{FigDrawing} (a). The noncollinear direction is due to the interaction between magnetic moments $\textbf{m}_i$ at $\textbf{r}_i$ and $\textbf{m}_j$ at $\textbf{r}_j$ as well as external magnetic field $\textbf{B}_\mathrm{ext}=\hat{\textbf{z}}B_\mathrm{ext}$ and magnetic anisotropy. 
\begin{align}
    E=& \sum_{ij}\Bigg(-J_\mathrm{ex}(\textbf{r}_{ij})\textbf{m}_i\cdot\textbf{m}_j+\textbf{D}(\textbf{r}_{ij})\cdot\left(\textbf{m}_i\times\textbf{m}_j\right)\notag\\
    & - \frac{\mu_0M^2}{4\pi}\frac{3\left(\textbf{m}_i\cdot\textbf{r}_{ij}\right)\left(\textbf{m}_j\cdot\textbf{r}_{ij}\right)-r^2_{ij}\textbf{m}_i\cdot\textbf{m}_j}{r^3_{ij}}\Bigg)\notag\\
    &-K\sum_j (m_{jz})^2-\sum_j M\textbf{m}_{j} \cdot \textbf{B}_\mathrm{ext}\notag\\
    \equiv&-\int d^3r\sum_j M_s\textbf{m}_{j} \cdot \textbf{B}_\mathrm{eff}.
\end{align}
Here $\textbf{r}_{ij}=\textbf{r}_i-\textbf{r}_j$. $M$ is the magnitude of the magnetic moment. $M_s$ is the saturation magnetization.  $K$ is the anisotropy constant. $\mu_0$ is the vacuum permeability. The total magnetic interaction includes dipole-dipole interaction, as well as the isotropic exchange interaction, governed by exchange constant $J_\mathrm{ex}$ and antisymmetric exchange interaction, governed by DMI vector $\textbf{D}$. At metallic interface, the exchange interaction can be mediated by itenerant conduction electrons (see~\ref{AppDMI} for detailed derivation).

The dynamics of $\textbf{m}_j$ is influenced by $\textbf{B}_\mathrm{eff}$ according to the Landau-Lifshitz-Gilbert (LLG) equation \cite{Gilbert2004}

\begin{align}
    \dot{\textbf{m}}_j=&-\gamma\textbf{m}_j\times\textbf{B}_\mathrm{eff}+\alpha \textbf{m}_j\times\dot{\textbf{m}}_j\notag\\
    =&-\gamma\textbf{m}_j\times\sum_i\Bigg[\textbf{B}_\mathrm{ext}+\left(\frac{J_\mathrm{ex}(\textbf{r}_{ij})}{M}-\frac{\mu_0M}{4\pi r^3_{ij}} \right)\textbf{m}_i\notag\\
    & -\frac{1}{M}\textbf{D}(\textbf{r}_{ij})\times\textbf{m}_i+\frac{\mu_0M}{4\pi r^3_{ij}}\textbf{r}_{ij}\left(\textbf{r}_{ij}\cdot\textbf{m}_i\right)\Bigg]\notag\\
    & -\frac{\gamma K}{M}\textbf{m}_j\times \hat{\textbf{z}}m_{jz}+\alpha \textbf{m}_j\times\dot{\textbf{m}}_j. \label{Eq.Larmor}
\end{align}
Here $\gamma$ is the gyromagnetic ratio, $\alpha$ is the dimensionless Gilbert damping and $M$ is magnitude of the magnetic moment. Since the interactions are short-ranged, the summation $\sum_j$ can be simplified to the nearest neighbors. The equation of motion can be simplified by the continuum limit of micromagnetic modeling.


Micromagnetic modeling (see Fig.~\ref{FigDrawing} for an illustration) focuses on the continuum limit of magnetic moments $\textbf{m}_i=\textbf{m}(\textbf{r}_i)$ and focuses on the unit direction $\textbf{m}$. 
Using Taylor expansion of $\textbf{m}(\textbf{r}_i)$ 
\begin{equation}
\textbf{m}(\textbf{r}_i)=\textbf{m}(\textbf{r}_j)+\textbf{r}_{ij}\cdot\nabla\textbf{m}(\textbf{r}_j)+\frac{1}{2}\left(\textbf{r}_{ij}\cdot\nabla\right)^2\textbf{m}(\textbf{r}_j)+\cdots .   
\end{equation}
and taking averaged over all direction of $\textbf{r}_{ij}$, with absolute value $|r_{ij}|=a$, one can simplify Eq.~\ref{Eq.Larmor} 
\begin{align}
    \dot{\textbf{m}}
    =&\gamma \textbf{m}\times\textbf{B}_\mathrm{ext} +\alpha  \textbf{m}\times\dot{\textbf{m}} +\frac{\gamma}{M_s} \textbf{m}\times\Bigg[K\hat{\textbf{z}}m_z-A\nabla^2\textbf{m}\notag\\
    & -C\nabla\left(\nabla\cdot\textbf{m}\right)+D_b\nabla\times\textbf{m}+D_i\left(\hat{\textbf{z}}\times\nabla\right)\times\textbf{m}\Bigg]\label{eqLLG}.
\end{align}
The reduced constants are
\begin{align*}
    A=&\left(J_\mathrm{ex}(a)-\frac{\mu_0 M^2}{4\pi a} \right)\frac{Na^2M_s}{2M}\approx J_\mathrm{ex}(a)\frac{Na^2M_s}{2M},\\
    C=&\frac{3\mu_0 M_sMNa}{16 },\\
    D_b=& \beta_b\frac{M_sm^2J^2N \left(1+\sin 2k_Fa\right)}{\hbar^4M},\\
    D_i=& \beta_i\frac{M_sm^2J^2N \left(1+\sin 2k_Fa\right)}{\hbar^4M}.
\end{align*}
$N$ is the number of nearest neighbor spins. $D_b$ ($D_i$) is the reduced DMI constant associated with the bulk (interfacial) spin-orbit coupling. For cobalt, $a=2.47$ \AA \cite{MaterialsProject}, $E_F=\hbar^2k_F^2/2m_e\approx 10$ eV \cite{C0CP00781A} and $H_{sd}\sim J/a \approx 0.13$ eV,
$A\approx 1.6\times 10^{-11}$ J/m$^2$ and $C=3.5\times 10^{-16}$ J/m$^2$.

Since $\textbf{m}$ is a unit vector, three components of $\textbf{m}$ in cylindrical coordinate 
\begin{equation}
    \textbf{m}=\hat{\textbf{r}}\sin\theta\cos\varphi+\hat{\boldsymbol{\phi}}\sin\theta\sin\varphi+\hat{\textbf{z}}\cos\theta
\end{equation}
can be described by polar $0\leq\theta<\pi$ and azimuthal angles $-\pi<\varphi\leq\pi$ . For magnetic skyrmion, which has a rotational symmetry, the polar angle can be assumed to have radial dependency $\theta(r,t)$. On the other hand, the azimuthal angle can be assumed to be uniform $\varphi(t)$ \cite{YU20211}. The value of $\varphi$ defines the helicity of the skyrmion. Skyrmion with $\varphi=0$ or $\pi$ has N\'eel-type helicity, while $\varphi=\pm \pi/2$ is the Bloch type. The equation of motion of the angles can be determined by manipulating Eq.~\ref{eqLLG} (see~\ref{AppTorque} for detailed derivation). 
\begin{align}
    &\left[\begin{array}{cc}
         1 & \alpha\sin\theta\\
         -\frac{\alpha}{\sin\theta}& 1 
        \end{array}\right]
        \left[\begin{array}{cc}
         \dot\theta\\
         \dot\varphi
        \end{array}\right]\notag\\
    &=\frac{\gamma}{M_s}
        \left[\begin{array}{cc}
         0\\
         M_sB_\mathrm{ext}+K\cos\theta 
        \end{array}\right]
        -\frac{\gamma A}{M_s\sin\theta}
        \left[\begin{array}{cc}
         0\\
         \frac{\theta'}{r}+\theta''-\frac{\sin\theta\cos\theta}{r^2}
        \end{array}\right]\notag\\
        &\ \ -\frac{\gamma C}{M_s} \cos\varphi\left(-\frac{1}{r^2}+\frac{\theta'\cot\theta}{r}-(\theta')^2\right)
        \left[\begin{array}{cc}
         \sin\theta\sin\varphi\\
         \cos\theta\cos\varphi
        \end{array}\right]\notag\\
        &\ \ -\frac{\gamma }{M_s}\sin\theta 
        \left[\begin{array}{cc}
         \theta'\left(D_b\cos\varphi+D_i\sin\varphi\right)\\
         \frac{1}{r}\left(D_b\sin\varphi-D_i\cos\varphi\right)
        \end{array}\right]
        . \label{EqConstLLG}
\end{align}

\section{Micromagnetic analysis of hybrid skyrmion}

The magnetic skyrmion can be obtained from the static solution $\dot{\theta}=0=\dot{\varphi}$ of Eq.~\ref{EqConstLLG}. Our objective is to find the conditions for the existence of hybrid skyrmion, which is a superposition of N\'eel and Bloch-type skyrmion. To determine the important ingredient for the existence of hybrid skyrmion, each of the following subsections \ref{seczeroDipol}, \ref{seczeroDbulk}, and \ref{seczeroDint} discuss cases for $C=0$ (no dipolar interaction), $D_b=0$ (no bulk DMI), and $D_i=0$ (no interfacial DMI), respectively. 

The discussions are complemented by a micromagnetic simulation using {\sc{Mumax3}} module \cite{Vansteenkiste2014}, which is used to analyze the dynamics of the skyrmion. However, the existing software can only implement bulk or interfacial DMI, but not both. Therefore, we perform micromagnetic simulations only for the cases $D_b=0$ or $D_i=0$. The module also does not accommodate the dipolar interaction setting, so we cannot run the simulation for $C=0$. However, it is naturally included since $\nabla\cdot \textbf{B}=0$ and therefore $C\neq0$. Common micromagnetic codes adopt the classical model of skyrmions in ultrathin multilayers that incorporate the dipolar interaction through an effective anisotropy term \cite{ivanov1990, rohart2013}. On the other hand, our analytical calculations are simpler than having to develop a code that can incorporate both $D_b$ and $D_i$ simultaneously.

\subsection{Zero dipolar interaction}
\label{seczeroDipol}
\begin{figure}[b]
    \centering
    \includegraphics[width=8.5cm]{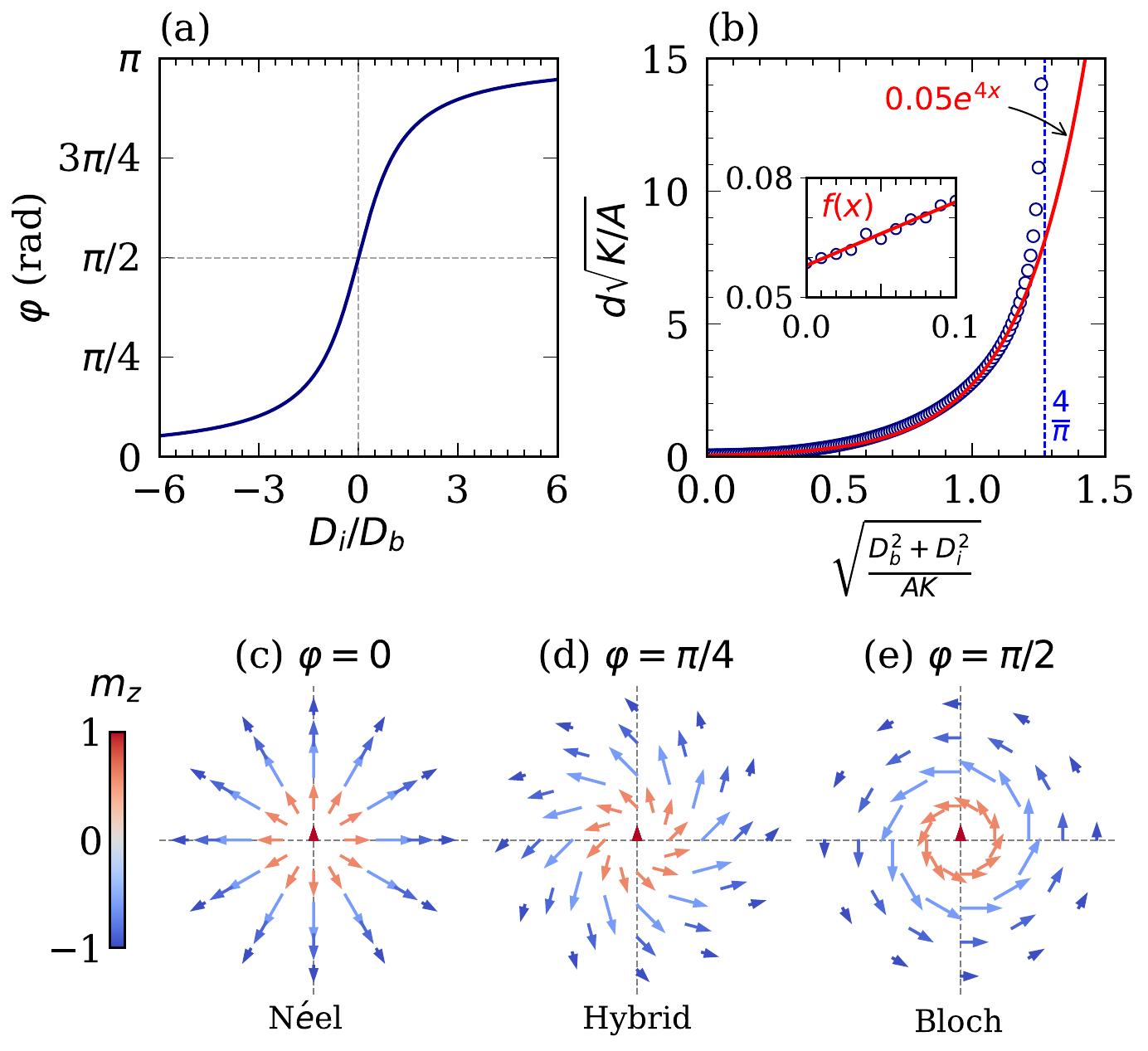}
    \caption{(a) Helicity $\varphi$ as a function of ratio of interfacial $D_i$ and bulk $D_b$ DMI for zero dipolar interaction $C=0$. (b) Diameter $d$ of skyrmion can be approximated by an exponential function, but diverges near the stability criteria $D_b^2+D_i^2=16/(\pi^2AK)$. For small value, it can be approximated with linear function $f(x)$. (c) $D_b=0,D_i=1.150\sqrt{AK}$ gives N\'eel skyrmion ($\varphi=0$), while (e) $D_i=0,D_b=1.150\sqrt{AK}$ gives Bloch skyrmion ($\varphi=\pi/2$). (d) $D_i=D_b=0.813\sqrt{AK}$ gives a hybrid skyrmion with $\varphi=\pi/4$, which is a superposition of Bloch and N\'eel skyrmion. }
    \label{fig:C0}
\end{figure}

At the stable state, both the polar and azimuthal angles remain constant over time. Examining Eq.~\ref{EqConstLLG} for $\dot\theta=0$, we can see the helicity of the hybrid skyrmion for zero dipolar interaction $(C=0)$ that is determined by the ratio of interfacial and bulk DMI strength ${D_i}/{D_b}$, as described in Fig.~\ref{fig:C0}(a),
\begin{align}
    0=& \theta'\left(D_b\cos\varphi+D_i\sin\varphi\right)\notag\\
    \varphi=& \frac{\pi}{2}+\arctan \frac{D_i}{D_b}.
\end{align}
One can control the helicity of hybrid skyrmion by tuning this ratio. For instance, the ratio ${D_i}/{D_b}=-1$ or ${D_i}={-D_b}$ will give a hybrid skyrmion with helicity $\varphi=\pi/4$. We note that ${D_i}/{D_b}$ is equal to the ratio of the interface and the bulk spin-orbit coupling strength, ${\beta_i}/{\beta_b}$.

Defining $R=r\sqrt{K/A}$, we can arrive at the differential equation for $\theta(r)$ by solving Eq.~\ref{EqConstLLG} in $\theta$ and $\varphi$ direction
\begin{align}
    &\frac{d^2\theta}{dR^2}+\frac{d\theta}{RdR}-\left(1+\frac{1}{R^2}\right)\sin\theta\cos\theta\notag\\
    &=\sqrt{\frac{D_b^2+D_i^2}{AK}}\frac{\sin^2\theta}{R} - \frac{M_sB_\mathrm{ext}\sin\theta }{K} .\label{Eq_R_C0}
\end{align}
Eq.~\ref{Eq_R_C0} can be solved numerically by assuming a small external field $B_\mathrm{eff}\ll K/M_s$, and setting boundary condition $\theta=0$ at origin and $\theta=\pi$ far from origin. Fig.~\ref{fig:C0}(b) illustrates the general trend of the diameter of the skyrmion as a function of the DMI strength, which can be approximated by an exponential function. For small DMI strengths, the diameter shows a linear dependence on $\sqrt{(D_b^2+D_i^2)/(AK)}$. The diameter diverges at $$\sqrt{\frac{D_b^2+D_i^2}{AK}}=1.273.$$ 
The value is in agreement with the stability criterion $D_{b,i}^2<(16/\pi^2)AK$ for pure N\'eel or Bloch skyrmion discussed in Ref.~\cite{Wang2018}. This result indicates that for $C=0$, the hybrid skyrmion requires a mixture of interfacial and bulk DMI and their ratio determines the helicity. In a practical system, no matter how small it is, $C$ is not zero. The next subsections will consider non-zero $C$ cases.

\subsection{Zero bulk Dzyaloshinskii-Moriya interaction}
\label{seczeroDbulk} 

In this subsection, we discuss the case of $D_b=0$. The helicity of the skyrmion can be determined from $\dot{\theta}=0$ 
\begin{align}
    0=&C\sin\theta\sin\varphi\cos\varphi\left(-\frac{1}{r^2}+\frac{\theta'\cot\theta}{r}-(\theta')^2\right)\notag\\
    &+\theta'D_i\sin\theta\sin\varphi 
\end{align}
This equation is fulfilled when $\sin\varphi=0$, which indicates a N\'eel skyrmion and
\begin{align}
    \cos\varphi =& \frac{D_i r^2\theta'}{C\left(1-r\theta'\cot\theta+r^2(\theta')^2\right)}\approx \frac{D_i}{D_0} \label{Eq.solusi2},
\end{align}
which indicates a hybrid skyrmion for $D_i\neq0$. 
The radius can be found from $\dot{\phi}=0$ 
\begin{align}
    &\frac{d^2\theta}{dR^2}+\frac{d\theta}{RdR}-\left(1+\frac{1}{R^2}\right)\sin\theta\cos\theta\notag\\
    &=\frac{D_i\cos\varphi}{\sqrt{AK}}\frac{\sin^2\theta}{R}+\mathcal{O}\left(\frac{M_sB_\mathrm{ext}}{K},\frac{C}{A}\right).\label{Eq_R_Db0}
\end{align}
Since Eq.~\ref{Eq_R_Db0} is similar to Eq.~\ref{Eq_R_C0}, one can see that the diameters 
\begin{align}
    d(\sin\varphi=0)=&\sqrt{\frac{A}{K}} f\left(\frac{D_i}{\sqrt{AK}}\right),\notag\\
    d\left(\cos\varphi=\frac{D_i}{D_0}\right)\approx&\sqrt{\frac{A}{K}} f\left(\frac{D_i^2}{D_0\sqrt{AK}}\right), \label{EqDiaDb0}
\end{align}
gives linear and quadratic dependency to $D_i$ for the N\'eel and hybrid skyrmions, respectively, where $f(x)$ is a linear function illustrates in the inset of Fig.~\ref{fig:C0}(b). Here, we use an approximation for Eq.~\ref{Eq.solusi2}. An exact solution requires a non-uniform of $\varphi$. 

\begin{figure}
    \centering
    \includegraphics[width=8.5cm]{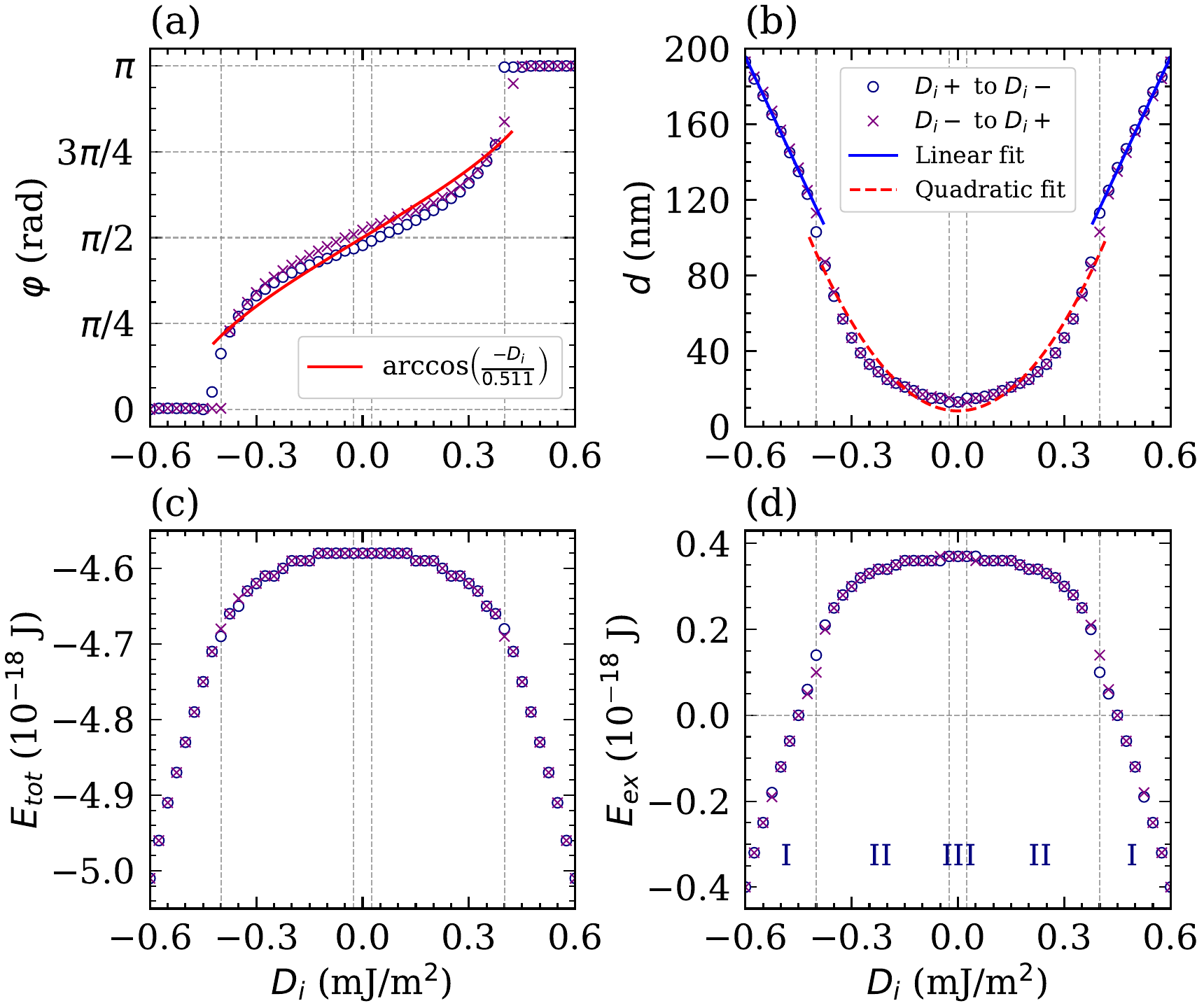}
    \caption{Micromagnetic simulation of (a) helicity, (b) diameter, (c) total energy, and (d) exchange energy of the skyrmion were examined against the changes of interfacial DMI. Linear dependency of the diameter is fit with $d=\pm 400~D_i - 44.33$, while the quadratic fit is $d=521.4~D_i^2 + 8.33$. The region I--III in the exchange energy graph correspond to the states of N\'eel, hybrid, and Bloch skyrmion, respectively.}
    \label{FigDint}
\end{figure}

The non-uniform treatment of $\varphi$ is implemented using micromagnetic simulation. We use {\sc{Mumax3}} to simulate an isolated skyrmion in a simple 512 nm square ferromagnetic lattice. We used a mesh size of 1 $\times$ 1 $\times$ 0.9 nm with a non-periodic boundary. Here, 0.9 nm is the ferromagnetic thickness. The simulation adopts Co-based magnetic parameters from \cite{helene2022,juge2019} which are $M_s = 1.42 \times 10^6$ A/m (saturation magnetization), $A = 16 \times 10^{-12}$ J/m (exchange stiffness), $K = 1.27 \times 10^6 $ J/m$^3$ (uniaxial anisotropy in the $z+$ direction) and $\alpha = 0.37$ (Gilbert damping parameter). { While Eq.~\ref{Eq_R_Db0} indicates that correction $B_\mathrm{ext}$ is in the order of $\mathcal{O}\left(M_sB_\mathrm{ext}/K\right)$, our simulation requires a non zero $B_\mathrm{ext}$ to stabilize the skyrmion. The value $B_\mathrm{ext}=4.6$ mT,  ($M_sB_\mathrm{ext}/K=5\times10^{-3}$), is chosen from Ref.~\cite{helene2022} that studies skyrmion for various values of $D_i$ with $D_b=0$ and $C=0$.}

Initially, we set the interfacial DMI constant $D_i = 0.6$ mJ/m$^2$ and place a N\'eel skyrmion at the center of the simulation region. We stabilize the system through the {\sc{Mumax3}} run function \cite{Mulkers2017}. This is to allow the system to reach minimum total energy. We then gradually reduce $D_i$ with a step of 0.025 mJ/m$^2$ until $D_i = -0.6$ mJ/m$^2$. At each stage, we let the skyrmion stabilize using the minimize function, which utilizes the steepest gradient method \cite{Exl2014}. Next, we run the same simulation, but with reversed values of $D_i = -0.6$ to $0.6$ mJ/m$^2$. For each step, we track the skyrmion diameter $d$, the helicity $\varphi$ (angle between the magnetic moment when $\theta=\pi/2$ and the radial direction), the exchange energy $E_\mathrm{ex}$, and the total energy $E_\mathrm{total}$.

\begin{figure*}
    \centering
    \includegraphics[clip,width=15cm]{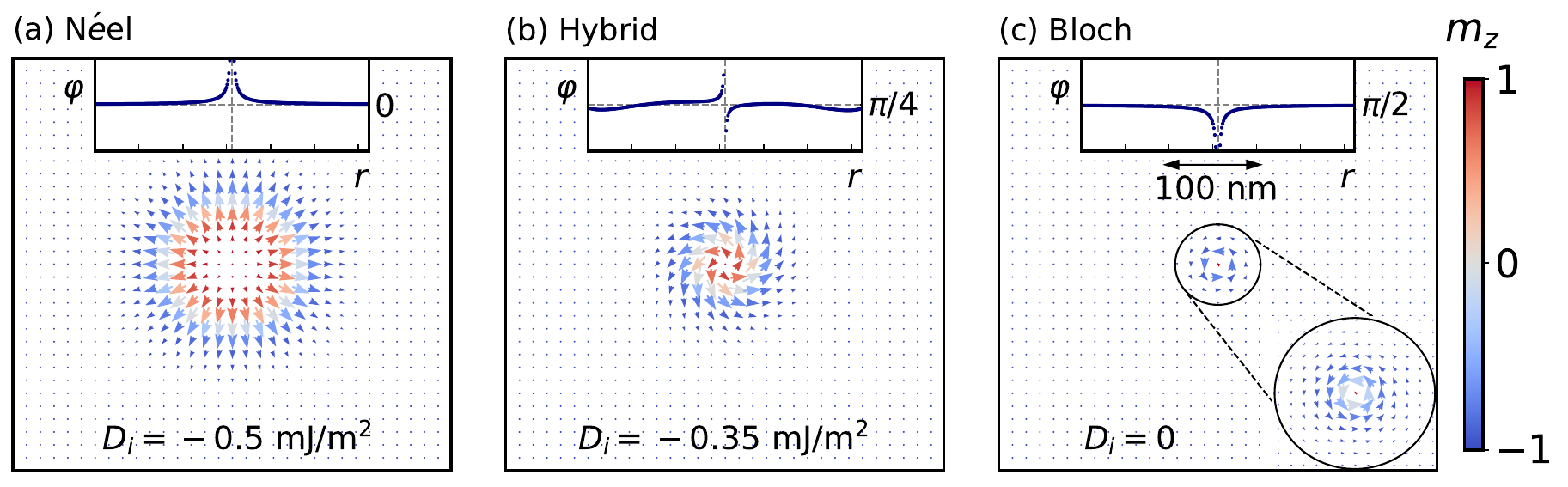}
    \caption{Micromagnetic simulation of stable states show the helicity transition between (a) N\'eel, (b) hybrid, and (c) Bloch skyrmions with varying value of interfacial DMI. The insets show the helicity as a function of position.}
    \label{FigSkyrm}
\end{figure*}

Fig.~\ref{FigDint} shows this simulation result. The first stabilization results in a N\'eel skyrmion with a diameter of 200 nm. Our micromagnetic simulation with given {magnetic parameters and DM constant} will give the same result regardless of whether we choose N\'eel or Bloch skyrmion  {provided by {\sc{Mumax3}}} as initial magnetization since we first stabilize the system using {\sc{Mumax3}} run function. If we choose the Bloch skyrmion as the initial magnetization for this case ($D_b=0$ and initial $D_i = 0.6$ mJ/m$^2$), it will eventually change into N\'eel skyrmion by the stabilization. The skyrmion remains in the N\'eel state until $D_i \approx 0.4$ mJ/m$^2$, then turns into a hybrid state. A small pure Bloch state ($\varphi=\pi/2$) with a diameter of 15 nm appears at a very narrow $D_i$ value around zero. When the value of $D_i$ is reached zero, the skyrmion diameter decreases and then increases again symmetrically, but with reverse chirality. In Fig.~\ref{FigDint}(a), the helicity changes gradually from 0 to $\pi$ with an antisymmetric trend and fits well with the function $\arccos$ as in Eq.~\ref{Eq.solusi2}. From the fitting, we can obtain the value of $D_0$ in Eq.~\ref{Eq.solusi2} to be 0.511 mJ/m$^2$.

Fig.~\ref{FigDint}(b) illustrates the linear and quadratic dependence of the diameters in Eq.~\ref{EqDiaDb0} for N\'eel and hybrid skyrmions, respectively. Fig.~\ref{FigDint}(a), we can arrive at the value of $D_0=0.511$ mJ/m$^2$. This agreement indicates that the uniform approximation of $\varphi$ is approximately true. Both simulations ($D_i+$ to $D_i-$ and $D_i-$ to $D_i+$) show a similar transition of N\'eel - hybrid - Bloch - hybrid - N\'eel skyrmion. Here, we found that in Co-based magnetic materials, a hybrid skyrmion appears within $0 < |D_i| < 0.4$ mJ/m$^2$. The helicity can be controlled only by adjusting the value of $D_i$. For example, $D_i=-0.35$ stabilizes a hybrid skyrmion with helicity $\varphi=\pi/4$. 

The plot of the total energy ($E_\mathrm{tot}$, Fig.~\ref{FigDint}(c)) and the exchange interaction energy ($E_\mathrm{ex}$, Fig.~\ref{FigDint}(d)) shows a symmetric pattern between the positive and negative $D_i$ constant. The trend appears as the inversion of the diameter change. 
The exchange energy plot (Fig.~\ref{FigDint}(d)) also shows three distinctive regions denoted by regions I to III. These regions are $|D_i| \geq 0.4$ mJ/m$^2$, $0 < |D_i| < 0.4$ mJ/m$^2$, and a very narrow area around $|D_i| = 0$ mJ/m$^2$. By correlating this with helicity, we can correspond region I to III with the skyrmion state of N\'eel, hybrid, and Bloch, respectively. Fig.~\ref{FigSkyrm} shows a stable state of N\'eel, hybrid, and Bloch skyrmion at $D_i= -0.5$, $-0.35$, and $0$ mJ/m$^2$, respectively. It also illustrates that $\varphi$ varies largely only near the center of the skyrmion and is relatively stable at distant locations.

\subsection{Zero interfacial Dzyaloshinskii-Moriya interaction}
\label{seczeroDint}

In this subsection, we discuss the case of $D_i=0$. The helicity of the skyrmion can be determined from $\dot{\theta}=0$ 
\begin{align}
    0 = &C\sin\theta\sin\varphi\cos\varphi\left(-\frac{1}{r^2}+\frac{\theta'\cot\theta}{r}-(\theta')^2\right)\notag\\
    &+\theta'D_b\sin\theta\cos\varphi \label{Eq.solusi3}
\end{align}
This equation is fulfilled when $\cos\varphi=0$, which indicates a Bloch-type skyrmion or
\begin{align}
    \sin\varphi =& \frac{D_b r^2\theta'}{C\left(1-r\theta'\cot\theta+r^2(\theta')^2\right)} \approx \frac{D_b}{D_0}. \label{Eq.solusi4}
\end{align}
Although the latter may indicate a hybrid skyrmion for $D_b\neq 0$, our micromagnetic simulation shows that only the former solution $\cos\varphi=0$ gives stable skyrmions. The radius of the stable Bloch skyrmion can be found by substituting $\cos\varphi=0$ to the solution of $\dot{\varphi}=0$ 
\begin{align}
    &\frac{d^2\theta}{dR^2}+\frac{d\theta}{RdR}-\left(1+\frac{1}{R^2}\right)\sin\theta\cos\theta\notag\\
    &=-\frac{D_b\sin\varphi}{\sqrt{AK}}\frac{\sin^2\theta}{R}+\mathcal{O}\left(\frac{M_sB_\mathrm{ext}}{K},\frac{C}{A}\right).\label{Eq_R_Di0}
\end{align}
The similarity of Eq.~\ref{Eq_R_Di0} to Eq.~\ref{Eq_R_C0} indicates a linear dependency of the diameter of skyrmion to $D_b$.

\begin{figure}[b]
    \centering
    \includegraphics[width=8.5cm]{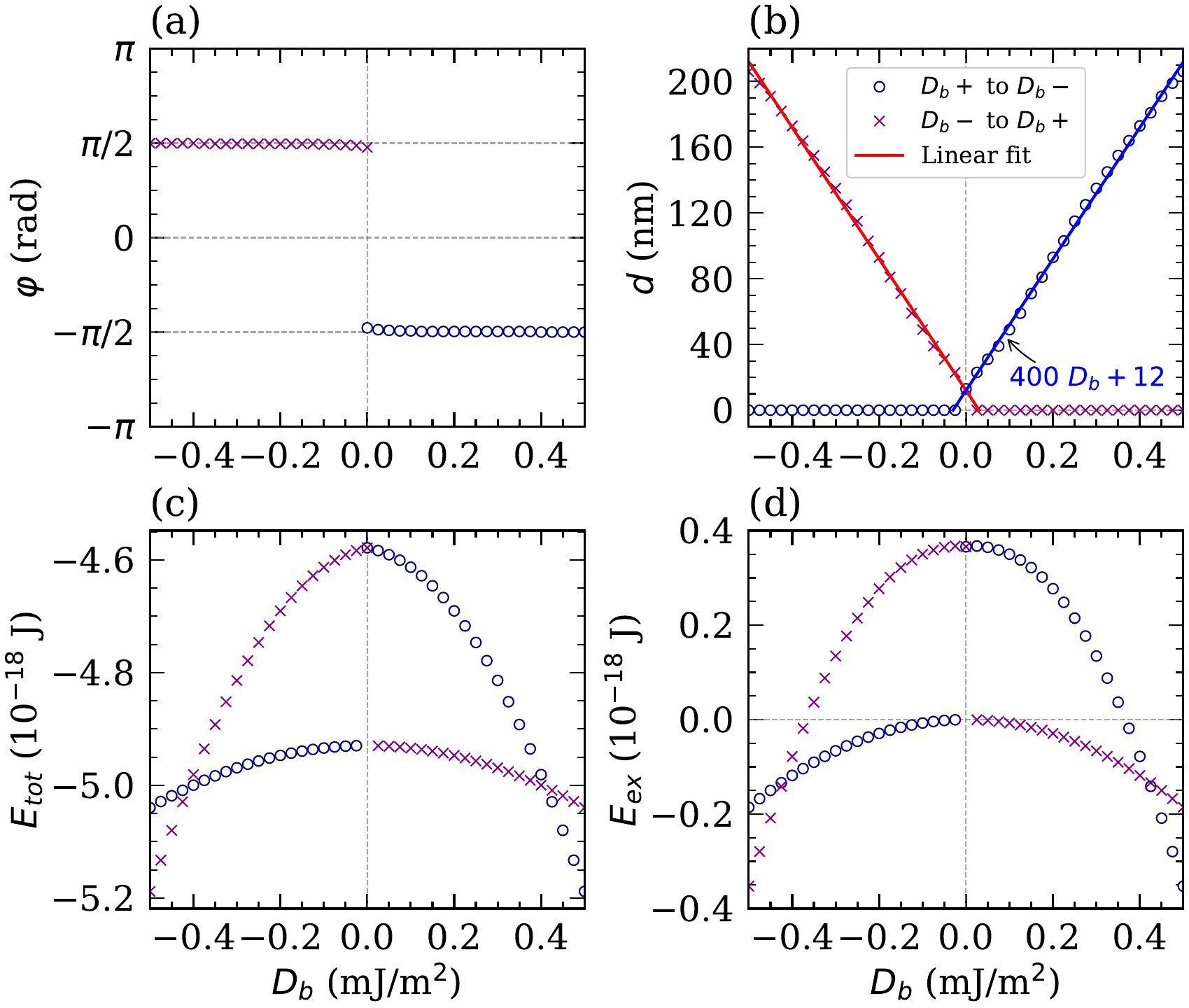}
    \caption{Micromagnetic simulation of (a) helicity, (b) diameter, (c) total energy, and (d) exchange energy of the skyrmion were examined against the changes of bulk DMI.}
    \label{FigDbulk}
\end{figure}

Fig.~\ref{FigDbulk} shows the result of our micromagnetic simulation. We simulate with a $D_b$ constant of $|D_b| \leq 0.5$. The magnetic parameter and the simulation procedure are the same as for the $D_b=0$ case. At the initial step, bulk DMI stabilizes a Bloch skyrmion with a diameter of around 200 nm. As the simulation progresses and $D_b$ decreases, the skyrmion remains in the Bloch state with a helicity of $\pi/2$ (Fig.~\ref{FigDbulk}(a)) and the diameter shrinks linearly (Fig.~\ref{FigDbulk}(b)). This is followed by Eq.~\ref{Eq_R_Di0}. The skyrmion vanishes (vanishing $d=0$ and divergent $\phi=|\inf|$) immediately after crossing $D_b = 0$ and does not recover along with the flip of the $D_b$ sign. The reversed simulation ($D_b-$ to $D_b+$) shows the same situation but with a helicity of $-\pi/2$. 

\begin{figure}
    \centering
    \includegraphics[width=8.5cm]{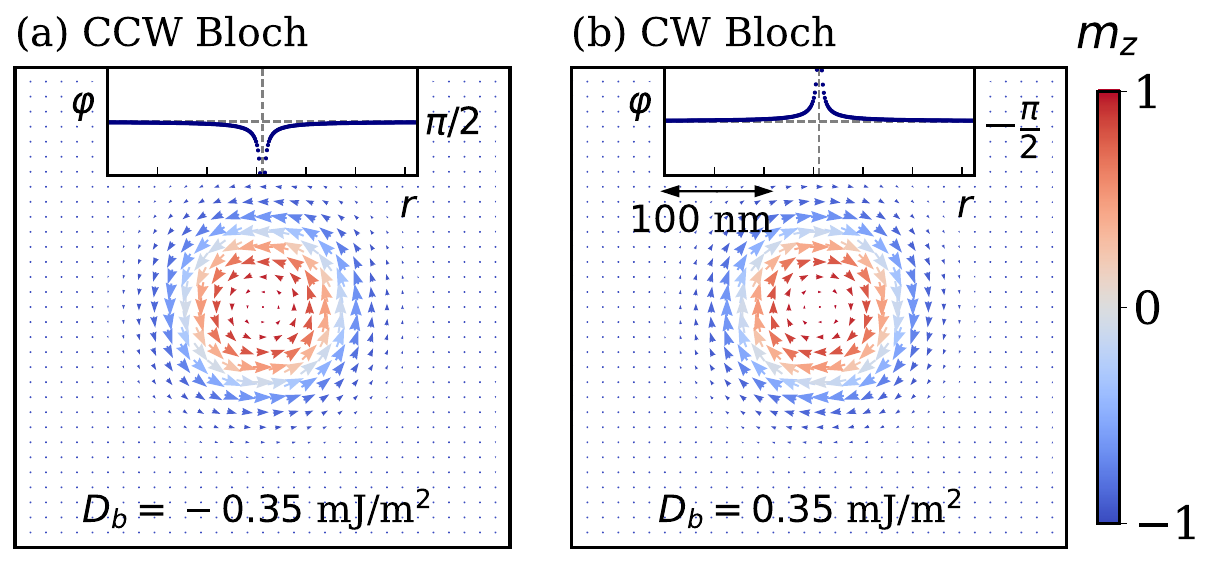}
    \caption{Micromagnetic simulation of stable states show that (a) counter-clockwise (CCW) Bloch skyrmion states occur when the $D_b$ are negative, while (b) clockwise (CW) Bloch skyrmion states are observed when the $D_b$ are positive.}
    \label{FigSkyrm-Bulk}
\end{figure}

The energy plot (Fig. \ref{FigDbulk}(c) and (d)) reflects this situation. After crossing $D_b = 0$, the energy drops as the skyrmion vanishes. The energy continues to decrease but at a slower rate as $D_b$ increases back to the opposite sign, but with the absence of the skyrmion. The energy at the final stage is still higher than that of the initial stage, which indicates that the bulk DMI with the existence of a skyrmion has a lower energy than one without a skyrmion. Here, we do not get any hybrid skyrmion in the absence of interfacial DMI. Fig.~\ref{FigSkyrm-Bulk} shows the stable Bloch skyrmion at $D_b=\pm 0.35$ mJ/m$^2$. The different $D_b$ sign gives different skyrmion chirality.

From these three simulation cases and with comparison to our analytical work, we find that analytically, all three cases can support hybrid skyrmion stabilization. However, our simulations showed a discrepancy that only the first two cases (no dipolar and no bulk DMI) stabilize hybrid skyrmion. When there is no interfacial DMI, although one of two analytical solutions can lead to hybrid skyrmion (Eq.\ref{Eq.solusi4}), the simulation confirms only the solution of Eq.\ref{Eq.solusi3} which leads to Bloch skyrmion. This emphasizes that hybrid skyrmion only exists with the existence of interfacial DMI ($D_i\neq0$) which indicates that the interfacial DMI is the main ingredient of hybrid skyrmion. It is the interplay between interfacial DMI with either dipolar or bulk DMI that will stabilize the hybrid magnetic skyrmion. It also implies that the dipolar interaction can stabilize the Bloch skyrmion since a small Bloch skyrmion exists in the total absence of all DMI ($D_i=D_b=0$, Fig.\ref{FigSkyrm} (c)). { Previous literature \cite{Fallon,Meijer,acsami.0c04661,adma.201807683} indicate that a stable hybrid skyrmion requires dipolar interaction between magnetic moments of different layers. In MuMax, skyrmion in a multilayer system requires a three dimensional modeling using a stack of two dimensional layers. However, our model uses a single layer system, therefore, the dipolar interaction is between magnetic moments at the same layer. This result shows the importance of intralayer dipolar interaction in stabilizing hybrid skyrmion.}

Our results also suggest that the helicity of the hybrid skyrmion can be controlled by two means. The first one is by tuning the ratio of $D_i/D_b$. Examples of systems containing both types of DMI are the crystals with $C_n$ symmetry classes, and a layered system with thin chiral magnet on the top of non-magnetic material with strong spin-orbit coupling.\cite{Kim2018} On the other hand, adjusting the $D_i$ constant will also tailor the helicity of the hybrid skyrmion when bulk DMI is absent. Since interfacial DMI exists mainly in heavy metal (HM) -- ferromagnet (FM) -- metal oxide (MO) layered system, tuning $D_i$ can be achieved experimentally by choosing the right heavy metal \cite{Ma2018, Jadaun2020}, altering FM thickness \cite{Lo2017, Cho2015}, or changing the oxidation state between the FM -- MO interface \cite{Diez2019, De2019}. Our calculation based on perturbation theory in section \ref{sec:theo} shows a relation between DMI with $k_F$ (Eq.~\ref{Eq6}) which is strongly affected by voltage-controllable Fermi energy. Therefore, voltage control can also tune the interfacial DMI as in ref.\cite{Suwardy2018, helene2022}.



\section{\label{sec:conc} Conclusions}

We have studied the stabilization of hybrid skyrmion in a two-dimensional system. There are three ingredients to stabilize hybrid skyrmion; interfacial DM, bulk DM, and dipolar interaction. The hybrid skyrmion can exist if $D_i \neq 0$. It is the interplay between interfacial DMI and either bulk DMI or dipolar interaction that stabilizes the hybrid skyrmion. Helicity can be controlled by tuning the ratio of ${D_i}/{D_b}$ in the absence of dipolar interaction, or simply by adjusting $D_i$ when $D_b$ is absent. In a Co-based magnet, we can find the hybrid skyrmion within $0 < |D_i| < 0.4$ mJ/m$^2$.  Our work implies that although micromagnetic simulation does not accommodate dipolar interaction, it is inherently included in the magnetic anisotropy term. We encourage the development of a micromagnetic code that enables the dipolar interaction setting to accommodate a system of material with non-unity relative permeability. 

\section*{Acknowledgements}
The authors acknowledge HPC-BRIN for computational resources. A.B.C. acknowledges support from Universitas Indonesia via PUTI Grant No. NKB-450/UN2.RST/HKP.05.00/2023. We acknowledge the BRIN visiting researcher fellowship in fiscal year 2023. Special thanks to H\'el\'ene B\'ea for valuable discussions.

\appendix

\section{Derivation of RKKY exchange constant and DMI vector}
\label{AppDMI}

At the ferromagnetic interface, there are localized and itinerant spins from unpaired $d$-electrons and conduction electrons, respectively \cite{Santiago2017}. The origin of itinerant spins of the conduction electron is spin polarization due to direct $s-d$ exchange interaction with localized spins \cite{CAHAYA2022168874}. 
\begin{equation}
    H_{sd}=-J\int d^2r  \boldsymbol{\sigma}\cdot\sum_n\textbf{m}_n\delta(\textbf{r}-\mathrm{r}_n),
\end{equation}
where $\textbf{m}_n$ is the direction of magnetic moment of a localized spin at $\textbf{r}_n$, $\boldsymbol{\sigma}=(\sigma_x,\sigma_y,\sigma_z)$ is the Pauli matrix, and $\psi(\textbf{r})=e^{i\textbf{k}\cdot \textbf{r}}$ is wave function of the conduction electron. 
In turn, itinerant spins can mediate indirect interactions between localized spins, known as the RKKY interaction \cite{RK,K,Y}, which can be described using perturbation theory on the non-perturbed energy of free-electron systems $E^{(0)}_k=\hbar^2k^2/2m$. 
\begin{align}
E_{\rm RKKY}=&\sum_{\textbf{p}\neq\textbf{k}} \frac{\left|\left<\textbf{p}\left|H_{sd}\right|\textbf{k}\right>\right|^2}{E^{(0)}_k-E^{(0)}_p}
\notag\\
\equiv&-\sum_{ij}J_{\rm ex}(\textbf{r}_i-\textbf{r}_j)\textbf{m}_i\cdot\textbf{m}_j.
\end{align}

\begin{figure}
    \centering
    \includegraphics[width=7.67cm]{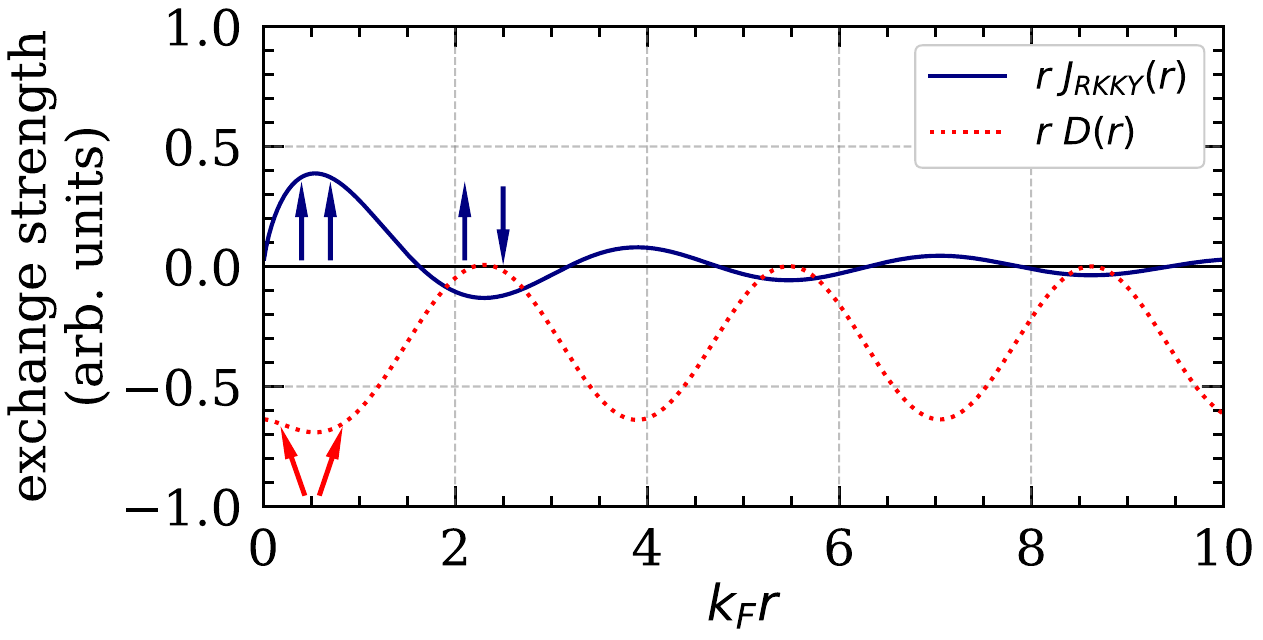}
    \caption{Exchange strength of symmetric RKKY interaction, $-J_\mathrm{RKKY}(|\textbf{r}_1-\textbf{r}_2|)\textbf{m}_1\cdot\textbf{m}_2$ (blue full line) and antisymmetric DMI, $-\textbf{D}(\textbf{r}_1-\textbf{r}_2)\cdot(\textbf{m}_1\times\textbf{m}_2)$ (red dotted line) as a function of distance $\textbf{r}_1-\textbf{r}_2$ between two magnetic moments.}
    \label{FigExchanges}
\end{figure}

Focusing on the magnetic interface, the exchange constant for the two dimensional system can be evaluated using cylindrical plane wave expansion \begin{equation*}
e^{i\textbf{k}
\cdot\textbf{r}}=\sum_n i^nJ_n(kr)e^{in
\theta},
\end{equation*}
where $J_n$ is the Bessel functions of the first kind. $J_{\rm ex}$ can be positive or negative-valued 
\begin{align}
    J_{\rm ex}=&\sum_{\textbf{k}}\sum_{\textbf{p}\neq\textbf{k}}  \frac{J^2e^{i(\textbf{p}\cdot\textbf{k})\cdot\textbf{r}}}{E^{(0)}_p-E^{(0)}_k}\notag\\
    =&\sum_\sigma\int \frac{d^2k}{(2\pi)^2}f_{k\sigma}\int \frac{d^2p}{(2\pi)^2} \frac{J^2e^{i(\textbf{p}\cdot\textbf{k})\cdot\textbf{r}}}{E^{(0)}_p-E^{(0)}_k}\notag\\
    =& \frac{4mJ^2}{\hbar^2}\int_0^{k_F} kdk \ J_0(kr) \int_0^{\infty} pdp \frac{J_0(pr)}{p^2-k^2} \notag\\
    =& \frac{4mJ^2}{\hbar^2}\int_0^{k_F} kdk \ J_0(kr)\frac{\pi}{2}Y_0(kr) \notag\\
    =&-\frac{\pi mJ^2}{\hbar^2}\left(J_0(k_Fr)Y_0(k_Fr)+J_1(k_Fr)Y_1(k_Fr)\right)\notag\\
    =&-\frac{3\pi mJ^2}{8\hbar^2}\frac{\sin 2k_Fr}{(k_Fr)^2}+\mathcal{O}(r^{-3}).
\end{align}
as illustrated in Fig.~\ref{FigExchanges}, which indicates ferromagnetic or antiferromagnetic coupling, respectively \cite{PhysRevB.58.3584}. Here $f_{k\sigma}=\theta(E_F-E)$ is the Fermi distribution. $E_F=\hbar^2k_F^2/2m$ is the Fermi energy.  $k_F$ is the Fermi wave-number, while $Y_n$ is the Bessel functions of the second kind. 

In systems with broken symmetry, such as magnetic interfaces, the energy of the conduction electron is strongly influenced by spin-orbit coupling. In Sec.~\ref{sec:DM}, the spin-orbit coupling is shown to be responsible for the DMI. In Sec.~\ref{sec:theo}, the interactions between localized spins are studied in the micromagnetic regime.

\subsection{\label{sec:DM} Dzyaloshinskii–Moriya interaction mediated by itinerant spin}

To describe the Dzyaloshinskii–Moriya interaction mediated by itinerant spin we consider the following Hamiltonian to describe the spin-orbit coupling \cite{10145485}
\begin{align}
H_\mathrm{so}=&\beta_i\boldsymbol{\sigma}\cdot\left(\hat{\textbf{z}}\times \textbf{k}\right)+\beta_b \boldsymbol{\sigma}\cdot\textbf{k},
\end{align}
where $\beta_b$ characterize the spin-orbit coupling due to non-centrosymmetric of the ferromagnetic bulk \cite{PhysRevB.96.115204,PhysRevX.5.011029}. $\beta_i$ is the coupling constant of the Rashba spin-orbit coupling due to broken symmetry at the interface \cite{Rashba1,Rashba2}. $\hat{\textbf{z}}$ is normal to the interface. The spin-orbit coupling of the system can induce an antisymmetric exchange, \textit{i.e.} DMI \cite{PhysRevLett.44.1538,PhysRevB.23.4667}. The influence of $H_\mathrm{so}$ on the indirect exchange interaction can be determined from the third perturbation terms
\begin{align}
E_{\rm DM}=&\sum_{\textbf{q}\neq\textbf{k}}\sum_{\textbf{p}\neq\textbf{k}} \frac{\left<\textbf{k}\left|H_1\right|\textbf{q}\right>\left<\textbf{q}\left|H_1\right|\textbf{p}\right>\left<\textbf{p}\left|H_1\right|\textbf{k}\right>}{\left(E^{(0)}_k-E^{(0)}_q\right)\left(E^{(0)}_k-E^{(0)}_p\right)}\notag\\
&-\left<\textbf{p}\left|H_1\right|\textbf{k}\right>\sum_{\textbf{p}\neq\textbf{k}} \frac{\left|\left<\textbf{p}\left|H_1\right|\textbf{k}\right>\right|^2}{\left(E^{(0)}_k-E^{(0)}_p\right)^2}\notag\\
=&\sum_{ij}\textbf{D}(\textbf{r}_i-\textbf{r}_j)\cdot\left(\textbf{m}_i\times\textbf{m}_j\right)
\end{align}
where the DMI vector can be evaluated using the following integral
\begin{align}
    \textbf{D}(\textbf{r})
    =&-iJ^2_\mathrm{ex}\sum_\textbf{k}\sum_{\textbf{p}\neq \textbf{k}}\frac{(\beta_i(\textbf{k}\times \hat{\textbf{z}}+\beta_b \textbf{k}))e^{i(\textbf{p}-\textbf{k})\cdot\textbf{r}}}{\left(E^{(0)}_p-E^{(0)}_k\right)^2}
    \end{align}
Since $\textbf{D}(-\textbf{r})=-\textbf{D}(\textbf{r})$, $\textbf{D}$ can be evaluated as follows
\begin{small}

    \begin{align}
    &\textbf{D}(\textbf{r})=\sum_\textbf{k}\sum_{\textbf{p}\neq \textbf{k}}\frac{i[\beta_i\hat{\textbf{z}}\times (\textbf{p}-\textbf{k}) +\beta_b (\textbf{p}-\textbf{k}))]J^2e^{i(\textbf{p}-\textbf{k})\cdot\textbf{r}}}{\left(E^{(0)}_p-E^{(0)}_k\right)^2}\notag\\
    &= (\beta_i\hat{\textbf{z}}\times\nabla+\beta_b \nabla)\sum_\sigma\int \frac{d^2k}{(2\pi)^2}f_{k\sigma}\int \frac{d^2p}{(2\pi)^2} \frac{J^2e^{i(\textbf{p}\cdot\textbf{k})\cdot\textbf{r}}}{\left(E^{(0)}_p-E^{(0)}_k\right)^2}\notag\\
    &= (\beta_i\hat{\textbf{z}}\times\nabla+\beta_b \nabla)\frac{8m^2J^2}{\hbar^4}\int_0^{k_F} kdk \ J_0(kr) \int_0^{\infty} pdp \frac{J_0(pr)}{p^2-k^2} \notag\\
    &= (\beta_i\hat{\textbf{z}}\times\nabla+\beta_b \nabla)\frac{4mJ^2}{\hbar^2}\int_0^{k_F} kdk \ J_0(kr)\frac{\pi r}{4k}Y_1(kr) \notag\\
    &= (\beta_i\hat{\textbf{z}}\times\hat{\textbf{r}}+\beta_b \hat{\textbf{r}})\frac{2\pi m^2 J^2k_F}{\hbar^4}J_0(k_Fr)Y_1(k_Fr)\label{Eq6}
\end{align}
\end{small}
Since
\begin{align}
    J_0(x)Y_1(x)=\left\{
    \begin{array}{cc}
       -\displaystyle\frac{2}{\pi x}  & x\ll 1 ,\\
       -\displaystyle\frac{1+\sin 2x}{\pi x}  & x\gg 1,
    \end{array} \right.\notag
\end{align}
the DMI vector is short-ranged, as illustrated in Fig.~\ref{FigExchanges}.

The total magnetic interaction can be found by including the dipolar interaction between localized magnetic moments, magnetic anisotropy, and Zeeman energy 
\begin{align}
    E=& \sum_{ij}\Bigg(-J_\mathrm{ex}(\textbf{r}_{ij})\textbf{m}_i\cdot\textbf{m}_j+\textbf{D}(\textbf{r}_{ij})\cdot\left(\textbf{m}_i\times\textbf{m}_j\right)\notag\\
    & - \frac{\mu_0M^2}{4\pi}\frac{3\left(\textbf{m}_i\cdot\textbf{r}_{ij}\right)\left(\textbf{m}_j\cdot\textbf{r}_{ij}\right)-r^2_{ij}\textbf{m}_i\cdot\textbf{m}_j}{r^3_{ij}}\Bigg)\notag\\
    &-K\sum_j (m_{jz})^2-\sum_j M\textbf{m}_{j} \cdot \textbf{B}_\mathrm{ext}\notag\\
    \equiv& -\int d^3r\sum_j M_s\textbf{m}_{j} \cdot \textbf{B}_\mathrm{eff}.
\end{align}
Here $\textbf{r}_{ij}=\textbf{r}_i-\textbf{r}_j$. 

\section{Constrained Landau-Lifshitz-Gilbert equation}
\label{AppTorque}
The dynamics of magnetization direction 
\begin{align}
    \textbf{m}=\left[\begin{array}{c}
         m_r\\
         m_\phi\\
         m_z
    \end{array}\right]
    =\left[\begin{array}{c}
         \sin\theta\cos\varphi\\
         \sin\theta\sin\varphi\\
         \cos\theta
    \end{array}\right]
\end{align} 
in polar coordinate can be reduced to the dynamics of the angles $\theta,\varphi$ by using $m_z$ and $m_+=m_r+im_\phi=e^{i\varphi}\sin\theta$.
\begin{small}
\begin{align}
    \frac{\partial}{\partial t}\left[\begin{array}{c}
         m_+\\
         m_z
    \end{array}\right]
    =\left[\begin{array}{cc}
         e^{i\varphi}\cos\theta & ie^{i\varphi}\sin\theta\\
         -\sin\theta & 0
    \end{array}\right]
    \left[\begin{array}{c}
         \dot\theta\\
         \dot\varphi
    \end{array}\right]
    =\left[\begin{array}{c}
         \tau_r+i\tau_\phi\\
         \tau_z
    \end{array}\right],    
\end{align}
\end{small}
where $\tau=\dot{\textbf{m}}$ is the torque from the LLG equation in Eq.~\ref{eqLLG}. Here, we limit the discussion to $\theta(r,t)$ and $\varphi(t)$.
$\tau$ consists of
    external magnetic field torque $\tau_\mathrm{ani}=-\gamma \textbf{m}\times \textbf{B}_\mathrm{ext}$, $\textbf{B}_\mathrm{ext}=\hat{\textbf{z}}B_\mathrm{ext}$
    \begin{align}
        \left[\begin{array}{c}
         \tau_+\\
         \tau_z
        \end{array}\right]_\mathrm{ext}
        =\left[\begin{array}{c}
         i\gamma B_\mathrm{ext}e^{i\varphi}\sin\theta\\
         0
        \end{array}\right],
    \end{align}
    magnetic anisotropy torque $\tau_\mathrm{ani}=-\gamma K\textbf{m}\times m_z\hat{\textbf{z}}/M_s$
    \begin{align}
        \left[\begin{array}{c}
         \tau_+\\
         \tau_z
        \end{array}\right]_\mathrm{ani}
        =\frac{\gamma K}{M_s}\left[\begin{array}{c}
         i e^{i\varphi}\sin\theta\cos\theta\\
         0
        \end{array}\right],
    \end{align}
    exchange torque $\tau_\mathrm{ex}=-\gamma A \textbf{m}\times\nabla^2\textbf{m}/M_s$
    \begin{align}
        \left[\begin{array}{c}
         \tau_+\\
         \tau_z
        \end{array}\right]_\mathrm{ex}
        =\frac{\gamma A}{M_s}\left[\begin{array}{c}
         -i e^{i\varphi}\left(\frac{1}{r}\frac{\partial\theta}{\partial r}+\frac{\partial^2\theta}{\partial r^2}-\frac{\sin\theta\cos\theta}{r^2}\right)\\
         0
        \end{array}\right],        
    \end{align}
    bulk DMI torque $\tau_b=\gamma D_b \textbf{m}\times\left(\nabla\times\textbf{m}\right)/M_s$
    \begin{small}
    \begin{align}
        \left[\begin{array}{c}
         \tau_+\\
         \tau_z
        \end{array}\right]_b
        =\frac{\gamma D_b}{M_s}\left[\begin{array}{c}
         -e^{i\varphi}\sin\theta\left(i\frac{\sin\theta\sin\varphi}{r}+\frac{\partial\theta}{\partial r}\cos\theta\cos\varphi\right)\\
         \frac{\partial\theta}{\partial r}\sin\theta\cos\varphi
        \end{array}\right]  ,      
    \end{align}
    \end{small}
    interfacial DMI torque $\tau_i=\gamma D_i \textbf{m}\times\left(\nabla m_z-\left(\nabla\cdot\textbf{m}\right)\hat{z}\right)/M_s$
    \begin{align}
        \left[\begin{array}{c}
         \tau_+\\
         \tau_z
        \end{array}\right]_i
        =\frac{\gamma D_i}{M_s}\left[\begin{array}{c}
         e^{i\varphi}\sin\theta\left(-i\frac{\sin\theta\cos\varphi}{r}+\frac{\partial\theta}{\partial r}\cos\theta\sin\varphi\right)\\
         -\frac{\partial\theta}{\partial r}\sin\theta\sin\varphi
        \end{array}\right]       , 
    \end{align}
    dipolar torque $\tau_\mathrm{dip}=-C\textbf{m}\times\nabla\left(\nabla\cdot\textbf{m}\right)/M_s$
    \begin{small}
    \begin{align}
        \left[\begin{array}{c}
         \tau_+\\
         \tau_z
        \end{array}\right]_\mathrm{dip}
        =&-\frac{\gamma C}{M_s}\cos\varphi\left(-\frac{\sin\theta}{r^2}+\frac{\cos\theta}{r}\frac{\partial\theta}{\partial r}-\left(\frac{\partial\theta}{\partial r}\right)^2\sin\theta\right)\notag\\
        &\cdot\left[\begin{array}{cc}
         i\cos\theta\\
         -\sin\theta\sin\varphi
        \end{array}\right],
    \end{align}
    \end{small}
    and damping torque $\tau=\alpha\textbf{m}\times\dot{\textbf{m}}$
    \begin{align}
        \left[\begin{array}{c}
         \tau_+\\
         \tau_z
        \end{array}\right]_\mathrm{dam}
        =\alpha\left[\begin{array}{cc}
         ie^{i\varphi} & -e^{i\varphi}\sin\theta\cos\theta\\
         0 & \sin^2\theta
        \end{array}\right]
        \left[\begin{array}{cc}
         \dot\theta\\
         \dot\varphi
        \end{array}\right].
    \end{align}
Combining all the torques, one arrive at Eq.~\ref{EqConstLLG}.

\section*{References}

\providecommand{\newblock}{}

\end{document}